\documentclass{article}

\usepackage{amssymb,amsmath}
\usepackage{fullpage}
\usepackage{graphicx}
\usepackage{appendix}
\usepackage{tikz}
\usepackage{amsmath}	
\usepackage{slashed}
\usepackage[margin=1.0in]{geometry}
\usepackage{setspace}
\usepackage{color}
\usepackage{fancyhdr}
\usepackage{collcell}
\usepackage{datatool}
\usepackage{environ}
\usepackage{latexsym}
\usepackage{amssymb}
\usepackage{epsfig,amsmath,graphics}
\usepackage{epstopdf}
\usepackage{verbatim}
\usepackage{wasysym}
\usepackage{feynmp-auto}
\usepackage{authblk}
\usepackage{xcolor}
\usepackage{enumitem}
\usepackage[utf8]{inputenc}
\usepackage{slashed}
\usepackage{cite}
\usepackage{setspace}
\begin{document}
\title{Low Frequency Gravitational Wave Detection with Gravito-Magnetism}

 \date{\today}
% \maketitle

\author[1,2]{Reza Ebadi}

\author[1]{Surjeet Rajendran}
\affil[1]{\small Department of Physics \& Astronomy, The Johns Hopkins University, Baltimore, MD  21218, USA}
\affil[2]{\small Department of Physics and Astronomy, University of Delaware, Newark, Delaware 19716, USA}
\maketitle

\begin{abstract}
The detection of gravitational waves in the low-frequency band (10 nHz to 1 micro-Hz) using inner solar system proof masses is severely limited by Newtonian gravity gradient noise (GGN) sourced by the asteroid belt. In this paper, we propose a novel detection concept that exploits the relativistic nature of gravitational waves to break the degeneracy between the signal and the asteroid GGN background. The non-relativistic motion of asteroids dominantly sources gravito-electric fields, with velocity suppressed gravito-magnetic components. Relativistic gravitational waves produce gravito-electric and gravito-magnetic components of equal magnitude. We propose detecting the gravito-magnetic component of low frequency gravitational waves using the Lorentz acceleration of test bodies. The measurement of this Lorentz acceleration is metrologically possible due to the fact that expected astrophysical sources of low frequency gravitational waves have considerably higher characteristic strains than the strains typically expected at higher frequencies. We show that this measurement allows for unambiguous extraction of the gravitational wave signal from the gravity gradient noise in a three satellite constellation.  The gravito-magnetic measurement still requires a proof mass that is well isolated from non-gravitational noise. We propose the use of  atom interferometers to serve as such a pristine inertial reference. 
\end{abstract}

\section{Introduction}
\label{sec:intro}
The historic detection of gravitational waves by the LIGO collaboration \cite{LIGOScientific:2016aoc} at frequencies above 10 Hz has ushered in a new era of astronomy. Just as the electromagnetic spectrum provides rich information about the universe across a variety of frequencies, there is an exceptionally strong case to probe the gravitational wave spectrum across a wide frequency range. Several observatories are currently under development to probe gravitational waves beyond LIGO. These include the LISA experiment at millihertz frequencies \cite{LISA:2017pwj},  mid-band detectors such as  MAGIS \cite{MAGIS-100:2021etm,Graham:2017pmn}, which aim to operate between the LISA and LIGO bands and next generation detectors in the LIGO band such as the Einstein Telescope \cite{Branchesi:2023mws}. Excitingly, remarkable experimental advances have also occurred at ultra-low frequencies: the NANOGrav collaboration \cite{NANOGrav:2023hvm} has recently reported strong evidence of a gravitational wave background at nanohertz (nHz) frequencies. Confirmation of this signal will strongly motivate efforts to detect gravitational waves in the ``low-frequency'' band (10 nHz to 1 $\mu$Hz). These NANOGrav measurements provide an empirical anchor to inform the sensitivities required to detect gravitational waves in this regime \cite{Sesana:2019vho}.

However, detecting gravitational waves in this frequency band is hampered by a significant source of background noise: the time-varying gravitational field (gravity gradient noise, or GGN) caused by the random motions of asteroids in the asteroid belt \cite{Fedderke:2020yfy}. Any detector relying on proof masses within the inner solar system would face acceleration noise from this asteroid motion 2--4 orders of magnitude larger than the target gravitational wave signal. This challenge has led to the suggestion that low-frequency gravitational wave detection requires distant proof masses. Indeed, pulsar timing arrays like NANOGrav circumvent this noise entirely because their proof masses (pulsars) are located far outside the solar system. However, the shot noise limitations of pulsar timing arrays necessitate a new class of distant proof masses, with the astrometric detection of non-magnetic white dwarfs emerging as a key contender \cite{Fedderke:2022kxq}. While this detection concept is promising, it faces daunting instrumental challenges: the instrument must be a Michelson stellar interferometer with a baseline in the 10--100 km range, operating at UV wavelengths of around 0.1 $\mu$m—parameters dictated by the emission spectra and luminosities of these stars. Although such an instrument might be possible, the technical difficulty of realizing this concept strongly motivates the development of alternative schemes to combat GGN in the low-frequency band.

In this paper, we propose exploiting the relativistic nature of gravitational waves to distinguish them from the GGN sourced by the non-relativistic motion of asteroids. The concept is theoretically straightforward and can be illustrated with an analogy in electromagnetism. Consider the detection of an electromagnetic wave in the presence of stray, non-relativistic moving charges. A relativistic electromagnetic wave possesses equal electric and magnetic field amplitudes. In contrast, the magnetic fields generated by a moving charge are suppressed by the charge's velocity. A sensor sensitive to magnetic fields can thus isolate the electromagnetic wave from the local charge background. This same principle applies to gravity. A relativistic gravitational wave has gravito-electric and gravito-magnetic components of equal amplitude, whereas the time-varying gravitational fields produced by moving masses dominantly source gravito-electric effects, with their gravito-magnetic components suppressed by the body's velocity. Thus, a detector sensitive to gravito-magnetic effects can effectively separate a true gravitational wave signal from the local GGN background.

How useful is this effect, and when can it actually be exploited? First, because the moving asteroids also create a velocity-suppressed gravito-magnetic effect, the maximum signal-to-noise suppression one can hope to achieve is set by the orbital velocity of the asteroid belt, $v \approx 10^{-4}$. This is a conservative estimate; while the asteroids share a bulk motion of $v \approx 10^{-4}$ around the Sun, the variance of the motion within this bulk flow is the actual source of the GGN. Because the expected gravitational wave signals are only 2--4 orders of magnitude weaker than the asteroid GGN, this conservative suppression factor is more than sufficient to permit signal extraction. Second, just as non-relativistic electromagnetic sensors are more sensitive to electric fields, non-relativistic sensors are dominantly sensitive to gravito-electric effects. Pursuing the suppressed gravito-magnetic signal is only viable if the intrinsic gravitational wave strains are large enough to lift it above the detector's noise floor. Fortunately, this is the case in the low-frequency band, where astrophysical sources are considerably brighter than those in higher-frequency bands like LIGO.

While gravito-magnetism permits the isolation of the gravitational wave signal from GGN, it does not intrinsically solve other metrological challenges, such as engineering proof masses sensitive solely to gravitational forces or achieving the requisite rotational stability for the measurement platform. We highlight how atom interferometers have the potential to provide proof masses with this necessary stability; however, a comprehensive analysis of such systems is left for future work. The primary focus of this paper is to establish the fundamental measurement protocol: given a sufficiently stable proof mass, gravito-magnetism enables inner solar system detectors to overcome asteroid GGN. This capability did not exist prior to our work, and the establishment of this measurement protocol is our primary goal.

The paper is organized as follows: Section~\ref{sec:signal} details how the gravito-magnetic signal manifests in an interferometer. Section~\ref{sec:protocol} outlines a realizable inner solar system detector configuration capable of extracting this signal. Section~\ref{sec:proofmass} describes how a proof mass with the required characteristics could be realized using atom interferometry, and we conclude in Section~\ref{sec:conclusions}.

\section{The Signal}
\label{sec:signal}

The physical effects of gravito-magnetism are similar to those of magnetism---magnetic fields ($\vec{B}$) cause precession of spins, and they also exert a velocity ($\vec{v}$)-dependent Lorentz force of the form $\vec{v} \times \vec{B}$ that is orthogonal to the direction of motion. Similarly, gravito-magnetism also causes the precession of spins (such as the rotation of the polarization of light) and also exerts a velocity-dependent force. The smallest gravitational wave characteristic strain $h_c$ that can be detected via spin precession is set by the shot noise limit $h_c \propto \frac{1}{\omega^2 T^2}\frac{1}{v_s\sqrt{N_s}}$, where $\omega$ is the frequency of the gravitational wave, $T$ is the interrogation time of the spin, $v_s$ is the relative velocity of the spin with respect to a local spin reference\footnote{The characteristic $\omega^2$ scaling arises due to the comparison with a local spin reference, since it is a Riemann effect. The frequency scaling can change if the spin is referenced to a distant reference. Note also that there is an additional precession $\propto h_c \omega^2 T L$, where $L$ is the separation between the spin and the spin reference.}, and $N_s$ is the total number of spins (whether polarized nuclear spins or stored photons) that are subject to the measurement during the course of the experiment. 

In the low-frequency band, the expected characteristic strains are of order $h_c \sim 10^{-15}$ to $10^{-16}$ \cite{Sesana:2019vho}, and we have not been able to find an experimentally realizable configuration of spins that can detect the expected precession. 

Instead, we focus on detecting the velocity-dependent Lorentz force acceleration on test bodies. As we will see, this offers a metrological path to the required sensitivity for two reasons. First, the measured acceleration scales with the baseline $L$ between different test bodies, and second, the ability to measure the relative acceleration in an interferometer depends on the wavelength $\lambda$ of the probe. The shot noise limit in this case scales as $h_c \propto \frac{\lambda}{L} \frac{1}{\omega^2 T^2} \frac{1}{v \sqrt{N}}$, where $v$ is the relative velocity between the test bodies, $T$ is the interrogation time of the test body, and $N$ is the total number of photons (or atoms) that are subject to the measurement during the course of the experiment. The boost provided by the ratio $\lambda/L$ makes the measurement of the Lorentz force metrologically sensitive enough to detect the gravitomagnetic effect\footnote{The comparison, though, is non-trivial---one can, for example, have relativistic spins in the form of polarized photons, boosting the spin precession signal relative to that of the Lorentz force. The ultimate comparison relies on technological limits on how long one can store photons/spins, as well as the ability to realize sufficiently fast and reliable proof masses for the measurement of the Lorentz force.}. To maximize the signal, we need interrogation times $T \sim \frac{1}{\omega}$, meaning the relative distance between the proof masses will  change by $v T \sim \frac{v}{\omega}$ during this time.

The above considerations necessitate a satellite experiment. In space, one can boost the signal by achieving a large baseline and relative velocity between test bodies. Further, space can also accommodate the large relative separation between the test bodies that is an inevitable consequence of having the large relative velocity and long interrogation time needed to see the low-frequency gravitational wave. Additionally, while gravito-magnetism can distinguish gravitational waves from gravity gradient noise, it still requires test masses that are dominantly sensitive to gravitational forces. While such proof masses can be realized terrestrially using freely falling atoms, the gravity of the Earth limits the free-fall time to at most $\sim 10$ s, far shorter than the period of the gravitational waves of interest.

\subsection{Setup}

\begin{figure}[htbp]
\centering
\begin{tikzpicture}[scale=1.2]
    % Sun
    \fill[yellow!60] (0,0) circle (0.6);
    \node at (0,0) {\textbf{Sun}};

    % Orbits (approximate)
    \draw[dashed, gray!50] (0,0) circle (3);
    \draw[dashed, gray!50] (0,0) circle (4.5);
    \draw[dashed, gray!50] (0,0) circle (5.5);

    % Satellites
    \node[circle, fill=black, inner sep=2pt, label=below:{$i$ (Master)}] (i) at (3,0) {};
    \node[circle, fill=black, inner sep=2pt, label=above:{$j$}] (j) at (45:4.5) {};
    \node[circle, fill=black, inner sep=2pt, label=right:{$k$}] (k) at (15:5.5) {};

    % Laser baselines
    \draw[->, thick, blue, dashed] (j) -- (k) node[midway, below, sloped] {\small $T, T+\tau, T+2\tau$};

     \draw[->, thick, blue, dashed] (i) -- (j) node[midway, above, sloped] {\small $\Delta^2\tau_{ij}$};

     \draw[->, thick, blue, dashed] (i) -- (k) node[midway, above, sloped] {\small $\Delta^2\tau_{ik}$};

\end{tikzpicture}
\caption{The three-satellite measurement constellation in solar orbit. Satellite $i$ serves as the master clock to synchronize the array. The deputy satellites exchange laser pulses at designated global times $T, T+\tau,$ and $T+2\tau$ to measure the discrete double time difference $\Delta^2\tau$. The multiple baselines are mathematically required to invert the signal matrix and simultaneously canceling higher-order gravity gradient noise from the asteroid belt.}
\label{fig:setup}
\end{figure}
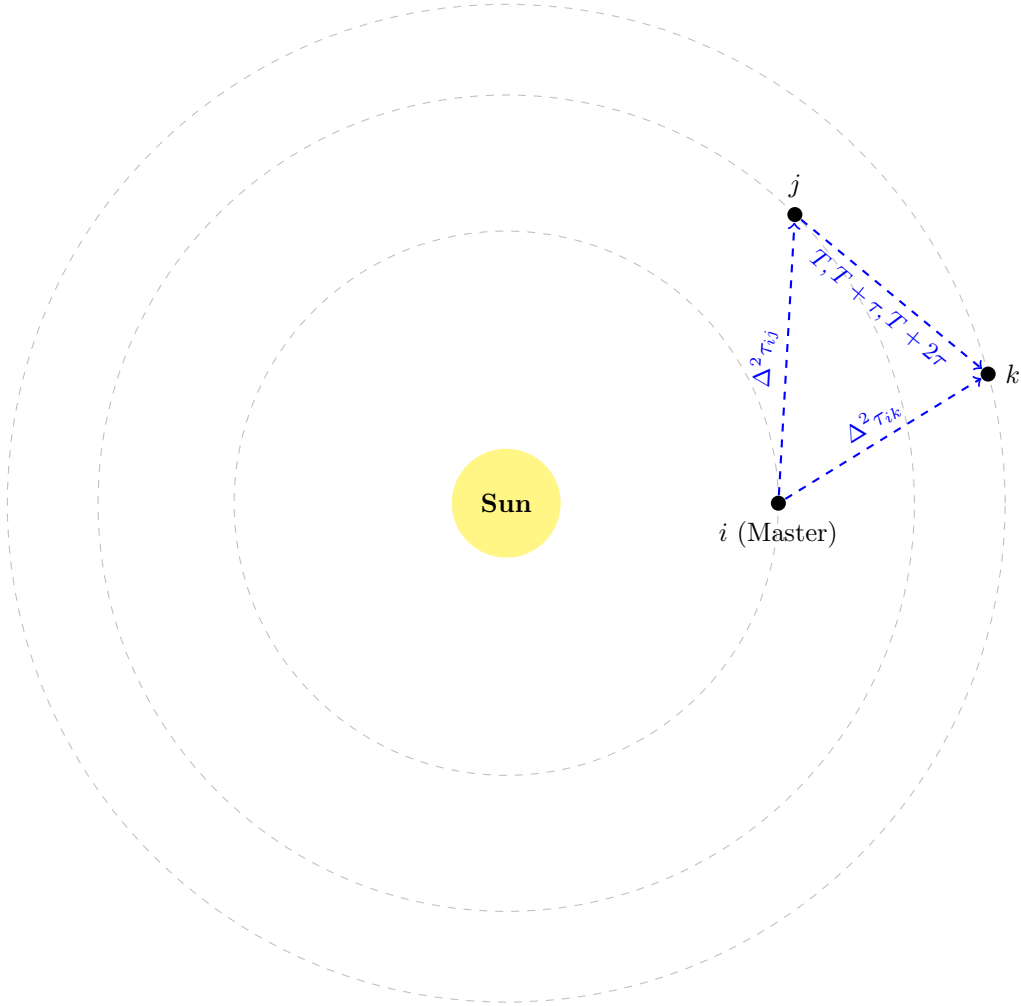

The basic experimental setup is as described in Figure \ref{fig:setup}. We consider three satellites labeled $i$, $j$ and $k$  that are in elliptical orbits around the Sun, with the orbits chosen so that they are within $\sim 10$ million km from each other. As we will see in section \ref{sec:protocol}, a three satellite configuration is sufficient to break the degeneracy between the gravitational wave signal and the gravity gradient noise.    The experiment will consist of different satellites sending pulses of light from one to another at pre-determined times and measuring the times of arrival of these pulses. While the emission and reception times of the pulses are measured using local clocks on the satellites, the experiment requires us to pick one of the satellites, say $i$, as the master clock and synchronize the local clocks of the other satellites with $i$ at an initial time $t = t_0$. This procedure is necessary since the satellites are moving with respect to each other and the gravitomagnetic effects of interest are terms that are linear in the velocity.  Corrections to the ticking rate of local clocks are second order in the velocity - but the surfaces of simultaneity are affected at linear order in the velocity. Thus when the satellites agree to send light at pre-determined times, they need to agree on a reference master time. 

We pick any satellite (say $j$) and send pulses of light from $j$ to one of the other satellites (say $k$) at the designated global times $T$, $T + \tau$ and $T + 2 \tau$ - that is $j$ uses the local clock on board $j$ to send these pulses, under the assumption that the initial synchronization between $i$ and $j$ has been unaffected. Gravitational waves and gravity gradients will change the inferred relation and this change is a (sub-dominant) source of the signal in this experiment. Satellite $k$ receives these pulses at times $T_{jk}\left(T\right), T_{jk}\left(T + \tau\right)$, and $T_{jk}\left(T + 2 \tau\right)$, as measured locally in satellite $k$'s frame of reference. Modulations in this arrival time are the dominant source of the gravitational wave signal and gravity gradient noise. To extract this, $k$ computes the relative acceleration of  $k$ with respect to $j$, isolating the gravitational signatures from  initial conditions (such as position and velocity). This is achieved by computing the discrete difference:

\begin{equation}
\Delta^2\tau_{jk} = T_{jk}\left(T + 2\tau\right) + T_{jk}\left(T\right) - 2 T_{jk}\left(T + \tau\right)
\label{eqn:doublediff}
\end{equation}

This basic term contains the gravity gradient and the gravitational wave signal. In the rest of this section, we focus on computing this term. While the signal and noise are present in just a single such term, the extraction of the gravitational wave from the gravity gradient, including the suppression of the latter to high enough order, requires additional baselines and thus the need for the three satellite configuration discussed in figure \ref{fig:setup}. These considerations are discussed in detail in section \ref{sec:protocol}. 

\subsection{The Calculation}
In this section, we focus exclusively on calculating the gravitational signals---the gravity gradient noise and the gravitational wave signal. Technical sources of noise, such as vibrations and laser frequency noise, can be added to these computations in a straightforward manner. 

To compute these gravitational signals, we need to pick a gauge to define the metric. It is conventional to describe the gravitational wave in the transverse-traceless (TT) gauge. In this gauge, a gravitational wave with the $+$ polarization moving along the $z$-direction (for example) with amplitude $h$ and frequency $\omega$ is described by the metric: 

\begin{equation}
ds^2 = -dt^2 + \left(1 + h \sin\left(\omega \left(t - z\right)\right)\right) dx^2 + \left( 1 - h \sin\left(\omega\left(t -z\right)\right)\right)dy^2 + dz^2
\end{equation}

The weak Newtonian gravitational fields that are the source of the gravity gradient noise can be described by a Newtonian potential $\Phi$. At the linearized level, this perturbation adds to the gravitational wave metric, yielding: 

\begin{equation}
ds^2 = - \left(1 + 2 \Phi\right) dt^2 + \left( 1 + h \sin\left(\omega\left(t - z\right) \right) - 2 \Phi\right) dx^2  + \left( 1 - h \sin\left(\omega\left(t - z\right) \right) - 2 \Phi\right) dy^2  + \left( 1  - 2 \Phi\right) dz^2 
\label{eqn:TTfull}
\end{equation}

While the metric is conventionally described in the above form, the physically measurable aspects of gravitation in a local experiment are most transparently described using Fermi normal coordinates. These are coordinates centered around the worldline of a freely falling observer, with the coordinate freedoms of gravity used to set the metric along this worldline to be exactly that of flat space and setting all the Christoffel symbols at this worldline to also be zero. Away from this worldline, the metric takes the form: 

\begin{equation}
g_{00} \approx -1 - R_{0a0b}\left(t_F\right)x^{a}x^{b}
\nonumber
\end{equation}
\begin{equation}
g_{0a} \approx -  \frac{2}{3}R_{0bac}\left(t_F\right)x^{b}x^{c}
\nonumber
\end{equation}
\begin{equation}
g_{ab} \approx \delta_{ab} - \frac{1}{3}R_{acbd}\left(t_F\right)x^c x^d
\label{eqn:FNC}
\end{equation}
where the Riemann tensor $R_{abcd}\left(t_F\right)$ is computed on the chosen worldline and $t_F$ is the time along this worldline. Notice that while the Riemann tensor itself can depend on the time $t_F$, the rest of the metric is solely a function of the spatial coordinates $x^{i}$. All of the time-varying effects of the gravitational wave (or, for that matter, any other time-dependent gravitational effect) arise solely through the time dependence of the Riemann tensor on the chosen worldline. 

In a realistic detector, the extraction of the gravitational wave follows standard techniques of Fourier analysis. We assume there is a measured time series. To check if the time series contains a gravitational wave at frequency $\omega$, we extract the Fourier coefficient of \eqref{eqn:doublediff} at $\omega$. The GGN background to this gravitational wave is the Fourier component of the GGN at $\omega$. For the rest of the analysis in this section, we assume that such a Fourier analysis has been performed and that we have reduced the problem down to separating the Fourier amplitude of the gravitational wave from the GGN. Since both of these oscillate in time the same way, we will drop the common $\cos\left(\omega t_F\right)$ and $\sin\left(\omega t_F\right)$ oscillations in these amplitudes when we write down our formulae below. Note that in the actual experiment, the proof masses will be in orbits around the Sun. Due to the modulation of the orbit, the gravitational wave and GGN terms will appear as sidebands around this carrier frequency. A specific data analysis protocol is necessary to extract the desired terms and we discuss these issues in section \ref{sec:protocol}.

For the problem at hand, we will develop the formalism to calculate \eqref{eqn:doublediff} and, without loss of generality, pick Fermi normal coordinates centered around satellite $i$. Let $\vec{\bar{x}}_{j,k}$ and $\vec{v}_{j,k}$ be the initial positions (at time $t_0$) and velocities of satellites $j$ and $k$ with respect to $i$. The accelerations of satellites $j$ and $k$ relative to $i$ are $\vec{a}_{j, k}$. The fundamental element from which the double differential $\Delta^2\tau_{jk}$ in \eqref{eqn:doublediff} is constructed is the quantity $T_{jk}\left(T\right)$, which is the time of arrival at satellite $k$, as measured by a local clock at satellite $k$, of light that was sent from satellite $j$ at the inferred global time $T$.  We are interested in this quantity to linear order in the speeds $|\vec{v}_{j,k}|$, the gravitational perturbations (both gravitational wave and gravity gradient), as well as the spatial separation between the satellites $j$ and $k$. At this order, the rate at which the clocks of $j$ and $k$ tick are the same as the ticking rate of $i's$ clock. Thus, times $T$ and $T_{jk}$ can be computed simply in the reference frame of the satellite $i$. 

\subsubsection{Unperturbed Paths}

To perform this computation, we first calculate the time taken by light to go from $j$ to $k$ in the absence of any acceleration. The trajectories of $j, k$ are: 

\begin{equation}
\vec{x}_{j, k}\left(t\right) = \bar{x}_{j,k} + \vec{v}_{j,k} \left(t - t_0\right)
\end{equation}

When light is sent from $j$ to $k$ at time $T$, the path of the light is: 

\begin{equation}
    \gamma\left(t\right) = \vec{x_{j}}\left(T\right) + \vec{V}_{\gamma}\left(t - T\right)
\end{equation}

Let us find the time $dT_{jk}$ when this light hits satellite $k$ to leading order in the velocities $v_{j,k}$. This computation also needs to establish the direction at which the light must be sent since the satellites have a transverse velocity relative to each other - that is, we need the actual velocity vector $\vec{V}_{\gamma}$ and not simply its magnitude. To find this, note that the condition for the light to hit satellite $k$ is: 

\begin{equation}
\gamma\left(T + dT_{jk}\right) = \vec{x}_k \left(T + dT_{jk}\right)
\end{equation}

yielding

\begin{equation}
\vec{x}_j\left(T\right) + \vec{V}_{\gamma} dT_{jk} = \vec{x}_k\left(T\right) + \vec{v}_k dT_{jk}
\label{eqn:lightdirection}
\end{equation}

Defining the vector: 
\begin{equation}
    \vec{L}_{jk}\left(T\right) = \vec{x}_k\left(T\right) - \vec{x}_j\left(T\right)
\end{equation}
we have: 

\begin{equation}
\vec{V}_{\gamma} = \frac{\vec{L}_{jk}}{dT_{jk}} + \vec{v}_k
\end{equation}
We still need to determine $dT_{jk}$. To do so, we demand that the light ray is a null vector and thus require the four velocity of the light to have zero norm. This translates into the requirement that $\vec{V}_{\gamma}.\vec{V}_{\gamma} = 1$. To leading order in $\vec{v}_k$, this implies: 
\begin{equation}
\frac{\vec{L}_{jk}.\vec{L}_{jk}}{dT_{jk}^2} + 2 \frac{\vec{L}_{jk}.\vec{v}_k}{dT_{jk}} = 1
\end{equation}
resulting in

\begin{equation}
dT_{jk} = \vec{L}_{jk}.\vec{v}_k + \sqrt{\vec{L}_{jk}.\vec{L}_{jk}}
\end{equation}

Define $\hat{n}_{jk} = \frac{\vec{L}_{jk}}{\sqrt{\vec{L}_{jk}.\vec{L}_{jk}}}$ to be the normal vector along the direction of $\vec{L}_{jk}$. With the above result for $dT_{jk}$, the velocity vector $\vec{V}_{\gamma}$ for the light is: 

\begin{equation}
\vec{V}_{\gamma} = \hat{n}_{jk} + \left(\vec{v}_k - \hat{n}_{jk} \left(\hat{n}_{jk}.\vec{v}_k\right)\right) = \hat{n}_{jk} + \vec{v}^{\perp}_{k}
\end{equation}
That is, the light is basically aimed at the direction of the unit vector from $j$ to $k$, but with a correction to its direction arising from the component $v^{\perp}_{k}$ of the velocity $\vec{v}_k$ that is perpendicular to the line of sight ({\it i.e.} the unit vector $\hat{n}_{jk}$). 

\subsubsection{Perturbed Paths}

We now add small accelerations $  \vec{a}_{j, \gamma}$ to the paths of the satellites and the light. We will work in the limit where we consider motions of the satellites occurring over time scales $ \omega T \lessapprox 1$. We also assume that the orbital frequency $f$ of the satellite is such that $f T \ll 1$. In this limit, the accelerations can be treated as constants and we can simplify the discussion of the algebra.  A more complete treatment can be performed using the full time dependent accelerations. It can be checked that this full treatment does not change the qualitative aspects of the signal and in fact yields the results of the following calculations in the limit of $  \omega T \lessapprox 1$. In this limit, the new positions of the satellites are: 

\begin{equation}
\vec{X}_j\left(t\right) = \vec{x}_j\left(t\right) +   \frac{\vec{a}_j}{2}\left(t - t_0\right)^2
\end{equation}

When $j$ sends light to $k$, the path of the light is:

\begin{equation}
\gamma\left(t\right) = \vec{x}_j\left(T\right) + \frac{1}{2}\vec{a}_j\left(T - t_0\right)^2 + \left( \vec{V}_{\gamma} +  \delta \vec{V}_{\gamma}\right) \left(t - T\right) +  \frac{\vec{a}_{\gamma}}{2}\left(t - T\right)^2
\end{equation}
where $\delta \vec{V}_{\gamma}$ accounts for the fact that the trajectory of the photon needs to be launched in a different direction in order for it to be received at $k$ in the presence of accelerations experienced by the photon and the satellite. For the light to still hit $k$, we need $\gamma\left(T + dT_{jk} +   \delta T_{jk}\right)$ to equal $\vec{X}_k\left(T + dT_{jk} + \delta T_{jk}\right)$. To linear order in the small acceleration, we have:

\begin{equation}
\frac{1}{2}\vec{a}_j\left(T -t_0\right)^2 + \vec{V}_{\gamma} \delta T_{jk} + \delta \vec{V}_{\gamma} dT_{jk} + \frac{\vec{a}_\gamma}{2}\left(dT_{jk}\right)^2 = \frac{1}{2}\vec{a}_k\left(T + dT_{jk} - t_0\right)^2 + \vec{v}_k \delta T_{jk}
\end{equation}
This allows us to obtain:

\begin{equation}
\delta \vec{V}_{\gamma} = \frac{1}{2 \, dT_{jk}}\left( \vec{a}_k\left(T + dT_{jk} - t_0\right)^2 - \vec{a}_j\left(T - t_0\right)^2\right)  - \frac{\vec{a}_{\gamma} dT_{jk}}{2} + \left(\vec{v}_k - \vec{V}_\gamma\right) \frac{\delta T_{jk}}{dT_{jk}}
\end{equation}
or using the expression for $\vec{V}_\gamma$

\begin{equation}
\delta \vec{V}_{\gamma} = \frac{1}{2 \, dT_{jk}}\left( \vec{a}_k\left(T + dT_{jk} - t_0\right)^2 - \vec{a}_j\left(T - t_0\right)^2\right)  - \frac{\vec{a}_{\gamma} dT_{jk}}{2} - \vec{L}_{jk}\frac{\delta T_{jk}}{dT_{jk}^2}
\end{equation}

To find $\delta T_{jk}$, we again enforce the condition that the light ray is null. The metric is of the form  $g_{\mu \nu} = \eta_{\mu \nu}+  \alpha_{\mu \nu}$, where $\alpha_{\mu \nu}$ is the small perturbation capturing the gravitational wave and the gravity gradient. In the co-ordinate system centered around $i$, the photon's trajectory in space-time is: 

\begin{equation}
\gamma_4\left(t\right) = \left(t, \vec{\gamma}\left(t\right)\right)
\end{equation}

Requiring that the norm of the photon's four-velocity is still zero, we have: 

\begin{equation}
\delta T_{jk} = \frac{1}{2}\frac{dT_{jk}}{\vec{V}_{\gamma}.\vec{L}_{jk}\left(T\right)}\left(\int_{T}^{T+dT_{jk}} \alpha_{\mu \nu} U^{\mu}U^{\nu} dt + \vec{V}_{\gamma}.\left(\vec{a}_k\left(T + dT_{jk} - t_0\right)^2  - \vec{a}_j\left(T - t_0\right)^2\right) \right)
\label{eqn:mainresult}
\end{equation}
where $U^{\mu}$ is the unperturbed four velocity of the light defined in terms of the time coordinate centered around $i$. Interestingly, observe that the acceleration $\vec{a}_{\gamma}$ of the light does not enter the answer. This is in fact a pleasing answer - it says that the time shift is the sum of the integral of the perturbed metric on the unperturbed paths of the light and a term that captures the motion of the boundaries (or satellites). This  form of the answer is a generic feature of the perturbed time (or phase) shift accrued in an interferometer when the probe particle obeys a classical equation of motion ({\it i.e.} a path that minimizes an action) in the absence of the perturbing accelerations. 

In the expression \eqref{eqn:mainresult} for the time shift $\delta T_{jk}$, observe that the integral term is parametrically smaller than the term that arises from the motion of the satellites. This is because in Fermi Normal coordinates, the deviations $\alpha_{\mu \nu}$ of the metric from flat space are $\propto [x]^2$ where $[x]$ symbolizes the spatial distance away from the origin $i$ of the Fermi-Normal coordinates. Thus the integral is proportional to $L^3$ where $L$ is the spatial separation between the satellites. The effects of the gravitational wave (and gravity gradient) that we are interested are the effects that are linear in the separation $L$. These arise from the boundary terms, where, as we will see below, the accelerations $\vec{a}_{j, k}$ are given by components of the Riemann tensor and are linear in the separation $[x]$ of $j, k$ from $i$. Moreover, in the experiment, when we take the discrete double difference \eqref{eqn:doublediff}, we will wait for times $\tau \sim \frac{1}{\omega}$ - for low frequencies, these times $\tau \gg L$ and thus the answer is always parametrically dominated by the motion of the boundaries\footnote{Note that the description of the physics in this language is gauge dependent - the above description captures the physics in Fermi Normal Coordinates. In the original transverse-traceless gauge, it can be checked that the gravitational wave signal arises from the perturbed integral, while the gravity gradient noise arises from the motion of the satellites. Ultimately, the only physical quantity is the gauge invariant time shift $\delta T_{jk}$ - but given a gauge, we can identify which terms are relevant for the calculation.}.

Thus, the time shift $\delta T_{jk}$ due to the acceleration is: 

\begin{equation}
\delta T_{jk} = \frac{1}{2} \frac{dT_{jk}}{\vec{V}_{\gamma}.\vec{L}_{jk}\left(T\right)} \left( \vec{V}_{\gamma}.\left( \vec{a}_k \left(T + dT_{jk} - t_0\right)^2 - \vec{a}_j \left(T - t_0\right)^2 \right) \right)
\label{eqn:minisigdeltaT}
\end{equation}

Substituting for $\vec{V}_{\gamma}$ and $d T_{jk}$, we have:

\begin{equation}
\delta T_{jk} = \frac{1}{2}\left( 1 + \hat{n}_{jk}.\vec{v}_{k}\right) \left( \hat{n}_{jk} + \vec{v}^{\perp}_{k}\right).\left( \vec{a}_{k}\left(T + dT_{jk} - t_0\right)^2 - \vec{a}_j\left(T - t_0\right)^2\right)
\label{eqn:deltaTjk}
\end{equation}

Thus, when $j$ sends light at time $T$ towards $k$, it is received by $k$ at time: 

\begin{equation}
T_{jk}\left(T\right) = T +  dT_{jk}  + \delta T_{jk}
\label{eqn:fullT}
\end{equation}

To use \eqref{eqn:deltaTjk}, we need to find the accelerations $\vec{a}_j$. In terms of the proper time $\tau_j$ of $j$, we have: 

\begin{equation}
\vec{a}_j = \frac{d^2 \vec{x}_j}{dt^2} = \frac{d \tau_j}{dt} \frac{d}{d \tau_j}\left( \frac{d \tau_j}{dt} \frac{d \vec{x}_j}{d\tau_j}\right) =\left( \frac{d\tau_j}{dt}\right)^2 \frac{d^2\vec{x_j}}{d\tau_j^2} + \frac{d\tau_j}{dt} \frac{d}{d\tau_j}\left(\frac{d\tau_j}{dt} \right) \frac{d \vec{x}_j}{d\tau_j}
\label{eqn:accdef}
\end{equation}

Since we are only interested in the accelerations to linear order in the gravitational fields, we can ignore the gravitational corrections to $\frac{d\tau_j}{dt}$ in the first term in \eqref{eqn:accdef}. The second term is non-zero only when $\frac{d}{d\tau_j} \left( \frac{d\tau_j}{dt}\right)$ is non-zero. This requires the derivative to either hit the position coordinates in \eqref{eqn:FNC} or the Riemann term. The second term thus contributes at the level of $\mathcal{O}\left(v^2\right)$ and at the level $\dot{R}_{abcd}[x]^2 v$. For a low frequency experiment in the solar system, the latter term is also $\mathcal{O}\left(v^2\right)$ or less. This can be seen by using orbital dynamics to estimate $ v \sim \omega [x]$ in concert with the fact that the orbital frequencies are comparable to the gravitational wave frequencies of interest. Both of these are parametrically smaller than the accelerations that we are interested in, which are linear in  the velocity and proportional to the Riemann tensor. Consequently, we have: 

\begin{equation}
    \vec{a}_j = \frac{d^2\vec{x}_j}{d\tau_j^2} 
\end{equation}
This can  be calculated using the geodesic equation for the metric \eqref{eqn:FNC}, yielding: 

\begin{equation}
\left(a_j\right)_{l} = -R_{0l0m}L^{m}_{ij} + 2 R_{lm0n}v^{m}L^{n}_{ij}
\label{eqn:accelerations}
\end{equation}
The velocity dependent, second term in \eqref{eqn:accelerations} arises from the $g_{0a}$ components of the metric in \eqref{eqn:FNC}. This term is the Lorentz-like acceleration caused by gravitomagnetism.

\subsubsection{The Gravitational Wave and Gravity Gradient}

To see how this term can distinguish between the gravity gradient and the gravitational wave, we write the metric \eqref{eqn:TTfull} in Fermi-Normal coordinates. The diagonal terms of the metric are: 

\begin{equation}
g_{00} = -1 - 2 \Psi -\frac{1}{2} h \omega^2\left(x^2 - y^2\right)\sin\left(\omega t_f\right)
\nonumber
\end{equation}

\begin{equation}
g_{xx} = 1 - 2 \Psi - \frac{1}{2}h \omega^2 z^2 \sin\left(\omega t_f\right)
\nonumber
\end{equation}

\begin{equation}
g_{yy} = 1 - 2 \Psi + \frac{1}{2}h \omega^2 z^2 \sin\left(\omega t_f\right)
\nonumber
\end{equation}
and 
\begin{equation}
g_{zz} = 1 - 2 \Psi 
\nonumber
\end{equation}
where $\Psi$ only contains the gravity gradient (and higher order) terms of $\Phi$ (the constant and linear terms are gauge transformed away in going to Fermi-Normal coordinates). Importantly, this metric also contains off-diagonal terms between time and space which evaluate to: 

\begin{equation}
g_{0x} = g_{x0} = -\frac{1}{2} h x z\omega^2 \sin\left(\omega t_f\right)
\nonumber
\end{equation}
\begin{equation}
g_{0y} = g_{y0} = \frac{1}{2} h y z\omega^2 \sin\left(\omega t_f\right)
\nonumber
\end{equation}
\begin{equation}
g_{0z} = g_{z0} = \frac{1}{4} h \left(x^2 - y^2\right)\omega^2 \sin\left(\omega t_f\right)
\label{eqn:FNCGW}
\end{equation}
These off-diagonal terms only depend on the gravitational wave. Comparing these terms to the general form of the Fermi-Normal coordinates in \eqref{eqn:FNC}, we see that the velocity dependent Lorentz acceleration is entirely sourced by these off diagonal gravitomagnetic terms. Measurement of this Lorentz acceleration will thus separate the gravitational wave signal from the gravity gradient noise. 

To get a feel for this answer, we consider the motion of satellites in x-z plane which we will take to lie along the ecliptic plane. For the metric \eqref{eqn:FNCGW}, we evaluate the acceleration of satellite $j$ to be: 

\begin{equation}
\vec{a}_j = -\nabla \Psi\left(\vec{x}_j\right) + h \omega^2 x^{1}_j\left(\left(-\frac{1}{2} + v^{3}_j\right) \hat{x} - v^{1}_j \hat{z}\right)
\label{eqn:netacceleration}
\end{equation}
where $x^{1}_j$ is the x-component of $\vec{x}_j$ and $v^{1,3}_j$ are the x and z components of $\vec{v}_j$.  In writing \eqref{eqn:netacceleration}, we have dropped the time dependent $\sin\left(\omega t_f\right)$ term, consistent with the approximation that we are interested in motion occurring at time scales $ \omega T \ll 1$ where we treat the accelerations to be approximately constant. At the level of the accelerations, the velocity dependent term breaks the degeneracy between the gravitational wave signal and the gravity gradient noise. But, the experiment does not directly measure the relative acceleration - instead, the acceleration is obtained from the discrete difference \eqref{eqn:doublediff} through the arrival times \eqref{eqn:deltaTjk}. The arrival time $\delta T_{jk}$ involves projection of the relative acceleration along velocity dependent vector directions and thus extraction of the velocity dependent parts of the acceleration is coupled to the ability to distinguish projections of the gravito-electric accelerations (of both the gravitational wave and the gravity gradient) along the velocity dependent directional vectors in \eqref{eqn:deltaTjk}. 

\subsubsection{Exploitable Features}

To disentangle the gravitational wave signal from the gravity gradient noise (GGN), it is instructive to consider a simple case where we assume that satellite $i$ is at the origin and satellite $j$ moves along a linear trajectory starting at $\left(L, 0\right)$ at $T = 0$. Its position as a function of the emission time $T$ is:

\begin{equation}
    \vec{x}_j(T) = \begin{pmatrix} L + v_x T \\ v_z T \end{pmatrix}
\end{equation}

For the following discussion, restrict the GGN potential to second order spatial gradients. 

\begin{equation}
    \Psi(\vec{x}) = \frac{1}{2}\Psi_{xx}x^2 + \Psi_{xz}xz + \frac{1}{2}\Psi_{zz}z^2
\end{equation}
These are the terms that are functionally degenerate with the gravitational wave signal and thus the most dangerous. We discuss disentangling higher order GGN corrections in section \ref{sec:protocol}. 

The resulting Newtonian acceleration $\vec{a}_{GGN} = -\nabla \Psi$ is:

\begin{align}
    a^{x}_{GGN}(T) &= -\Psi_{xx}(L + v^{(j)}_x T) - \Psi_{xz}(v^{(j)}_z T) \\
    a^{z}_{GGN}(T) &= -\Psi_{xz}(L + v^{(j)}_x T) - \Psi_{zz}(v^{(j)}_z T)
\end{align}

The effective line-of-sight acceleration is projected via the kinematic vector $\vec{N} = (1 + \hat{n} \cdot \vec{v})(\hat{n} + \vec{v}^\perp)$. To leading order in velocity, with $\hat{n} \approx \hat{x}$ and $\vec{v}^\perp \approx v^{(j)}_z \hat{z}$, this vector expands as $\vec{N} \approx (1 + v^{(j)}_x)(\hat{x} + v^{(j)}_z \hat{z}) \approx \hat{x} + v^{(j)}_x \hat{x} + v^{(j)}_z \hat{z}$. The projected acceleration $\vec{N} \cdot \vec{a} \approx a^x + v^{(j)}_x a^x + v^{(j)}_z a^z$ for the GGN is therefore:

\begin{equation}
    \vec{N} \cdot \vec{a}_{GGN} \approx \left[ -\Psi_{xx}(L + v^{(j)}_x T) - \Psi_{xz}(v^{(j)}_z T) \right] + v^{(j)}_x \left[ -\Psi_{xx} L \right] + v^{(j)}_z \left[ - \Psi_{xz} L \right]
\end{equation}

where we have dropped terms of $\mathcal{O}(v^2)$. Rearranging this into terms that depend on $L$ and $v T$ yields:

\begin{equation}
    \vec{N} \cdot \vec{a}_{GGN} \approx \underbrace{-\Psi_{xx}L - \Psi_{xx}v^{(j)}_x L - \Psi_{xz} v^{(j)}_z L}_{\text{L}} \quad \underbrace{- \Psi_{xx} v^{(j)}_x T - \Psi_{xz} v^{(j)}_z T}_{\text{v T}}
\end{equation}

Similarly, projecting the acceleration due to the gravitational wave along the line of sight yields: 

\begin{equation}
    \vec{N} \cdot \vec{a}_{GW} \approx -\frac{1}{2}h\omega^2 L - \frac{1}{2}h\omega^2 v^{(j)}_x L + h\omega^2 v^{(j)}_z L
\end{equation}

Comparing the constant terms of these projections reveals a dangerous degeneracy. The GGN projection contains the velocity-coupled term $-\Psi_{xz} v^{(j)}_z L$, which arises from the transverse velocity dotting with the transverse Newtonian acceleration $a^z$. This offset is parametrically identical to the gravitomagnetic term in the gravitational wave projection, $h \omega^2 v^{(j)}_z L$. However, the GGN  term is intrinsically accompanied by the resolving terms, $-\Psi_{xx} v^{(j)}_x T - \Psi_{xz} v^{(j)}_z T$, which arise because the motion of the satellite ($x = v^{(j)}_x T, z = v^{(j)}_z T$) implies a changing longitudinal acceleration along the x-axis. 

Combining these projections, the total perturbed time shift $\delta T_{ij}(T) \propto \frac{1}{2} (\vec{N} \cdot \vec{a}) T^2$ becomes a cubic polynomial in $T$:

\begin{equation}
    \delta T_{ij}(T) = \frac{1}{2} C_0^{(j)} T^2 + \frac{1}{2} C_1^{(j)} T^3
\end{equation}

where the coefficients are:

\begin{align}
    C_0^{(j)} &= -\Psi_{xx}L - \Psi_{xx}v^{(j)}_x L - \Psi_{xz}v^{(j)}_z L - \frac{1}{2}h\omega^2 L - \frac{1}{2}h\omega^2 v^{(j)}_x L + h\omega^2 v^{(j)}_z L \\
    C_1^{(j)} &= -\Psi_{xx} v^{(j)}_x - \Psi_{xz}v^{(j)}_z
\end{align}

Taking the discrete double difference \eqref{eqn:doublediff}, we get the observable: 

\begin{equation}
    \Delta^2  \tau_{ij}(T) = C_0^{(j)} \tau^2 + C_1^{(j)} \left(3 \tau^2 T + 3 \tau^3\right)
    \label{eqn:finalsig}
\end{equation}
 By evaluating this discrete double difference for different values of the delay parameter $\tau$ and initial times $T$\footnote{Note that the initial time $T$ explicitly appears in this expression since it sets the position  of $j$ at the start of the experiment.}, we can definitively measure the time-dependent coefficient $C_1^{(j)} = -\Psi_{xx} v^{(j)}_x - \Psi_{xz} v^{(j)}_z$ for satellite $j$. However, this single measurement contains two unknowns, $\Psi_{xx}$ and $\Psi_{xz}$. To disentangle them, we utilize a second baseline to a satellite $k$ on a different trajectory, yielding $C_1^{(k)} = -\Psi_{xx} v^{(k)}_x - \Psi_{xz} v^{(k)}_z$. Because the velocity vectors of $j$ and $k$ are distinct, this forms an invertible linear system that perfectly resolves both $\Psi_{xx}$ and $\Psi_{xz}$. Once these gradients are isolated using multiple baselines, they can be subtracted from $C_0^{(j)}$, cleanly breaking the degeneracy and isolating the gravitational wave signal.

 \subsubsection{Algebraic Extraction}
 \label{subsec:algebra}

An additional signal extraction protocol is also possible with some algebraic pre-processing of the data stream. Observe that in \eqref{eqn:fullT}, the times $T$ and $dT_{jk}$ are known - they are the known times at which the signal is sent and the light travel time taken to go from $j$ to $k$ when the motion is inertial. When $k$ receives a pulse from $j$, $k$ can subtract these known contributions. The datastream of $k$ is thus solely just $\delta T_{jk}$, containing only the acceleration signal. Next, since the pre-factor $\left( 1 + \hat{n}_{jk}.\vec{v}_k \right)$ is also known, this pre-factor can be divided out to yield the data-stream: 

\begin{equation}
    P_{jk} = \left(\hat{n}_{jk} + \vec{v}^{\perp}_k\right).\left(\vec{a}_k - \vec{a}_j \right)\left(T-t_0\right)^2
\end{equation}
where we have ignored the $dT_{jk}$ light travel time correction to the displacement, since it is higher order in the relative distance $L$. 

To see how we can use $P_{jk}$ to extract the accelerations, let us focus on $P_{ij}$ and $P_{ji}$ (where $i$ is the origin): 

\begin{equation}
    P_{ij} = \left(\hat{n}_{ij} + \vec{v}^{\perp}_j\right).\left(\vec{a}_j  \right)\left(T-t_0\right)^2
\end{equation}

\begin{equation}
    P_{ji} = \left(-\hat{n}_{ij} \right).\left(-\vec{a}_j  \right)\left(T-t_0\right)^2
\end{equation}

The difference $P_{ij} - P_{ji}$ is $\vec{v}^{\perp}_j.\vec{a}_j\left(T - t_0\right)^2$. Since we know $\vec{v}^{\perp}_j$, we can use $P_{ij} - P_{ji}$  to extract the component of $\vec{a}_j$ that is perpendicular to $\hat{n}_{ij}$. Once this is known, we can use the sum $P_{ij} + P_{ji}$ to extract the component that is along $\hat{n}_{ij}$. This allows us to thus fully reconstruct the time series $\vec{a}_j \left(T - t_0\right)^2$. Similarly, we can also extract  $\vec{a}_k \left(T - t_0\right)^2$. In these accelerations, there is no degeneracy between the gravity gradient and the gravitomagnetic effects of the gravitational wave. 

\section{Measurement Protocol}
\label{sec:protocol}

The measurement of the discrete double difference $\Delta^2 \tau$ along two baselines $ij$ and $ik$ with three satellites is sufficient to disentangle the gravitational wave signal from the leading gravity gradient noise terms $\Psi_{xx}, \Psi_{xz}$ and $\Psi_{zz}$. In the absence of the gravitomagnetic signal, these terms are functionally degenerate with the gravitational wave signal and are thus the biggest hindrance to gravitational wave detection. However, the gravity gradient is a Taylor series and thus it contains terms of arbitrarily high order. While the higher order terms are functionally different from the gravitational wave signal, they will contribute to the measurement of $\Delta^2 \tau$ and they need to be algebraically removed from the data stream. Due to the functional independence between these terms and the gravitational wave signal, this algebraic process is possible, provided there are sufficiently many measurements available to subtract out GGN terms of the required order.

The higher order GGN terms are rapidly suppressed when the distance between the satellites decreases. But, the signal in the setup also decreases when the distance between the satellites is decreased. This is both due to the standard linear suppression of the Riemann acceleration of the gravitational wave signal and an additional linear suppression of the relative velocity between satellites in orbit when their orbital parameters are close to each other. There is thus a tension between the need to suppress the gravity gradient noise and have the metrological capability to detect the signal. The latter depends upon the technology used for the measurement, including the ability of this technology to suppress other technical sources of noise like vibrational motion. We imagine using an optical interferometer to measure the distance between two proof masses. As discussed in section \ref{sec:proofmass}, the proof masses will be stabilized by local atom interferometers in each satellite to achieve the desired stability. The demands on the stability of the proof mass are set by the relative acceleration caused by the gravitomagnetic signal in the setup, which is boosted by the baseline between the satellites. But, a larger baseline implies that higher order gravity gradient signals will also contribute to the measurement. These will then need to be distinguished from the gravitomagnetic acceleration. As we will see in section \ref{sec:proofmass}, we need LISA-like baselines $\sim 10 \times 10^6$ km so that the relative acceleration from the gravito-magnetic signal is large enough to be detectable by atom interferometers. As we will see below, such a baseline requires measurement and subsequent cancellation of gravity gradients up to the seventh order (where the second order GGN is the standard term $\Psi_{xx}$). It is possible to achieve this cancellation with three satellites in elliptical orbits with $\mathcal{O}\left(1\right)$ loss in sensitivity.

\subsection{Metrology and the Gravity Gradient}
\label{subsec:metrology}
The fundamental observable in this experiment is the discrete double time difference \eqref{eqn:finalsig}. The gravitomagnetic time shift is $\sim h L v \omega^2 \tau^2$ where $L$ is the average separation between the satellites and $v$ the relative (transverse) velocity. If the satellites are in a solar orbit of radius $R \sim 1$ AU, their relative velocities are $v \sim \sqrt{\frac{GM_\odot}{R}} \frac{L}{R} \sim 10^{-4} \frac{L}{R}$. We can imagine measuring this time shift using an optical interferometer, resulting in a phase shift $\delta \phi \sim 10^{-4} k h L^2 \omega^2 \tau^2/R $, where $k$ is the wave-number of the light. For a mirror of waist $w_0$, the Rayleigh range is $z_R =  k w_0^2/2$ and  the number of photons received at the distance $L$ is $N \sim \frac{P}{k} \left(\frac{z_R}{L} \right)^2$, where $P$ is the input power.   Taking $\tau \sim 1/\omega$, the shot noise sensitivity is: 

\begin{equation}
h \approx 2 \times 10^4 \frac{R}{L} \frac{1}{\sqrt{k^3 P}} \frac{1}{w_0^2}\approx \frac{10^{-13}}{\sqrt{\text{Hz}}}\left( \frac{10 \times 10^6 \text{ km}}{L}\right) \sqrt{\frac{\mu\text{W}}{P}}  \left(\frac{\text{0.3 m}}{w_0}\right)^2 \left(\frac{\text{eV}}{k}\right)^{\frac{3}{2}}
\label{eqn:SNRCon}
\end{equation}
which is at the level required for gravitational wave detection in the low frequency (10 nHz - 100 nHz) band. The fact that input powers as low as $\sim \mu$W is sufficient to detect the velocity suppressed gravitomagnetic signal is fundamentally due to the large expected characteristic strains of low frequency signals, in contrast to the metrological problem encountered at higher frequency gravitational wave detection. The limit on the baseline $L \sim 10^7$ km is not set by the optical metrology - but rather by the difficulty of obtaining suitably stable proof masses, as discussed in section \ref{sec:proofmass}.  

The gravity gradient noise from the asteroid belt $\Psi_{xx}$ is 100 - $10^4$ times the expected gravito-electric signal $\sim h L \omega^2$. That is, if $R_A \sim 3$ AU is the distance to the asteroid belt, the gravity gradients $\Psi_{xx}, \Psi_{xz}, \Psi_{zz}$ are order $\Psi_{xx} L \approx 10^4 h L \omega^2$. Since the gravitomagnetic signal is smaller by $\sim 10^{-4} \frac{L}{R}$, the required shot noise sensitivity is high enough to be sensitive to higher order gravitational gradients such as $\Psi_{xxx}$ which would give rise to a signal $\Psi_{xxx} L^2 \sim \left(\Psi_{xx} L\right) \frac{L}{R_A}$. The highest order GGN that the experiment is sensitive to is set by equating $\Psi_{xx} L \left(\frac{L}{R_A}\right)^n \lessapprox 10^{-4} h \frac{L^2}{R} \omega^2$. For $L \sim 10$ million km, this equates to $n = 6$. The shot noise is thus sensitive to the seventh order terms in the GGN. These terms will have to be distinguished from the gravitational wave signal.

\subsection{Signal Correction from Higher Order GGN}
The derivation of \eqref{eqn:deltaTjk} focused on terms linear in the baseline $L$, as this is the leading order at which the gravitational wave strain $h$ and the second-order Gravity Gradient Noise (GGN) couple to the time-delay observable. Higher-order GGN gradients scale with higher powers of the baseline ($L^2, L^3, \dots$). How do we distinguish these functionally different terms from the gravitational wave signal?  In fact, there is danger lurking at the level of the third order GGN signal which scales as $L^2$. This is because the relative velocity between satellites in an orbit scales as $L$ and thus the gravitomagnetic signal does in fact scale as $L v \propto L^2$. Despite this fact, the third order GGN is functionally distinct from the gravitomagnetic signal since its temporal signature is distinct. Much like the acceleration from the second order GGN, the acceleration from the third order GGN is proportional to $\Psi_{xxx} \left(L + v T\right)^2$ - there is thus always a temporal contribution to this acceleration arising from the physical motion of the satellites through the GGN. Such a temporal contribution does not exist for the gravitomagnetic acceleration and it is thus possible to distinguish the gravitational wave signal from the GGN. 

The problem then reduces to ensuring that the constellation provides a sufficient number of independent measurements to span the degrees of freedom of these higher-order terms, allowing the isolation of the gravitational wave signal from the GGN. Let us now track the different points in our derivation where we dropped terms that were higher order in $L$ and identify which of these will dominate the corrections from higher order GGN. Note that we are interested in terms that would ruin the $\mathcal{O}\left(v\right)$ degeneracy breaking between the gravitomagnetic effects and the GGN. Effects that are $\mathcal{O}\left(v^2\right)$ can thus be neglected.

We discuss the terms that we dropped and their relevance to the expanded GGN analysis. 

\begin{enumerate}

\item Higher order GGN contributes to the acceleration vector $\vec{a}_j = -\nabla \Psi $. We need  to distinguish these terms from the signal.  Notice that the higher order corrections here are powers of $L/R_A$ and $v T/R_A$. In the low frequency regime, for satellites in solar orbits where the gravitational wave period ($\sim T$) is comparable to the orbital frequency, the distance $v T$ travelled by the satellites is $\sim L$. Thus the higher order GGN terms will be polynomials in $L$ and $v T$ and their distinct polynomial form in comparison to the gravitational wave signal will permit us to mathematically distinguish the signal from the higher order GGN. We thus keep all the $v T$ corrections to $\nabla \Psi$. 
\item In \eqref{eqn:mainresult}, we ignored the integral and focused on the motion of the end-points. This approximation is still valid. The accelerations receive higher order GGN corrections. The position change caused by them scales as $\sim T^2 \nabla \Psi$, which in the double differential becomes a term $\sim \nabla \Psi$. The integral produces an effect that is $\mathcal{O}\left(L^3\right)$ in the light travel time. Since $L$ changes by $v T$, the double differential produces an effect $\propto \mathcal{O}\left(v^2\right)$ which can be safely neglected. 
\item In \eqref{eqn:accdef}, we approximated the co-ordinate acceleration $\vec{a}_j$ by the geodesic acceleration. This approximation is still valid since the higher corrections change the acceleration at $\mathcal{O}\left(v^2\right)$. 
\item The second term where the higher order GGN manifests itself is through the change in the ticking rate of the local clock at the locations of the satellites $j$ (the emitter) and the $k$ (the receiver). To leading order in the perturbation,  these correct the terms $T + \Delta T_{ij}$ in  equation \eqref{eqn:fullT}. The dominant change comes from the change in the emission time $T_{j,k} \gg L$ at which the clocks emit and receive the pulses. We keep track of these changes. There is a smaller change that arises due to changes in the definitions of the instantaneous lengths $L_{jk}$. These terms start out being $\mathcal{O}\left(L^3\right)$ - with one power of $L$ arising from the definition of $L_{jk}$ and the other two arising from the fact that the clock shift is $\mathcal{O}\left(L^2\right)$. Much like the case of the integral that we neglected above, this term also contributes as $v^2$ in the double differential and can be safely neglected.  
\end{enumerate}

Incorporating the above, we see that the time-shift \eqref{eqn:fullT} receives a correction to $\vec{a}_j$ from the higher order GGN terms and a change to the emission times themselves. The latter are calculated to be:

\begin{equation}
\Delta T_{jk}\left(T\right) = \int_{t_0}^{T + dT_{jk}} dt\left( \left( \Psi\left(\vec{x}_k\left(t\right)\right) + \left(\frac{h \omega^2 \left( x^{1}_k\left(t\right)\right) ^2}{4}\right) \right) \right)-  \int_{t_0}^{T} dt \,\left(\left( \Psi\left(\vec{x}_j\left(t\right)\right) + \left(\frac{h \omega^2 \left( x^{1}_j\left(t\right)\right) ^2}{4}\right) \right) \right)
\end{equation}

Thus the total signal is:

\begin{equation}
s_{jk}\left(T\right) = T_{jk}\left(T\right) + \Delta T_{jk}\left(T\right)
\end{equation}

We then find the discrete derivatives: 

\begin{equation}
\Delta^{2} \tau_s = s_{jk}\left(T + 2 \tau\right) + s_{jk}\left(T\right) - 2 s_{jk}\left(T + \tau\right)
\label{eqn:GGNGW}
\end{equation}
for various values of $\tau$ and baselines and ask how many baselines and measurements are needed to distinguish the higher order GGN from the gravitational wave signal. 

It can be checked that the gravitational wave signal can be distinguished from the GGN even with three satellites and three baselines. But, a key issue in such an extraction is the question of the signal-to-noise penalty that one may have to suffer while extracting a signal that is nearly degenerate with the noise. To understand the issue, let us consider a toy problem. Suppose we had a measurement data stream for the form $\frac{G\mu}{R^3}\delta x + h R \omega^2$. We can in principle distinguish the signal $h$ from the noise parameter $\delta x$ if we measured this data stream for two different values of $R$ at say $R_{1,2}$. Mathematically, as long as $R_{1} \neq R_{2}$, the signal $h$ can be distinguished from the noise $\delta x$. However,  we expect that as $R_2 \rightarrow R_1$, it should get harder to distinguish the two. This intuition is indeed correct - we have to equate this data stream to some measurement apparatus which is ultimately limited by some shot noise $\epsilon$. It can be checked that for a given noise floor $\epsilon$, the sensitivity to $h$ scales $\propto \epsilon \left(R_2 - R_1\right)^{-1}$. This is because fundamentally we are trying to solve the matrix problem $M s = \epsilon$ where $s$ is a vector containing the signal $h$ and the noise $\delta x$. $M$ represents the geometric matrix that contains information about the parameters $R_{1,2}$.  When $R_2$ approaches $R_1$, the matrix $M$ contains degenerate rows and thus its inverse $M^{-1}$ becomes singular. The signal extraction problem is solved by $s = M^{-1}\,\epsilon$. Note that $\epsilon$ contains contributions from the $h$, $\delta x$ and shot noise. The matrix $M^{-1}$ is mathematically defined to perfectly invert and solve for $h$ and $\delta x$. Thus, even when as $M^{-1}$ becomes singular, it retains the capability to solve for $h$ and $\delta x$. However, when it multiplies the random shot noise terms, the singular values amplify the shot noise resulting in reduced sensitivity to the signal. This can thus be viewed as a SNR penalty that must be paid to distingish $h$ from $\delta x$ - that is, the shot noise would need to be lowered in order to extract the signal from the shot noise background. 

In this particular measurement strategy, we know that we already have to pay a SNR penalty - that is, the gravitational wave signal and the GGN are degenerate up to the gravitomagnetic signal which is suppressed by the relative velocity $v$. Thus, we know that we need a sensor whose shot noise is smaller by a factor of $v$ relative to a detector that measures the gravito-electric effect of the gravitational wave. Moreover, distinguishing the higher order GGN relies fundamentally on the satellites moving around in their orbit and covering displacements $v T \sim L$ during the course of their orbit. Thus, a SNR penalty that is $\sim v$ is inevitable and this is already factored into our discussions in sub-section \ref{subsec:metrology} and section \ref{sec:proofmass}. However, since the required SNR capabilities are already at the state-of-the-art, we do not want to pay additional suppressions. 

To analyze these issues, we considered three cases. In all of these cases, we take a central satellite that is in orbit around the Sun at a period $\sim$ year. We then consider various numbers of deputy satellites that are also in solar orbits, but with their orbits near the central satellite. The distances between the deputies and the central satellite was restricted to be at most $\sim 10^7$ km.  The orbits are taken to be actual solutions to the solar system dynamics - for the case of elliptical orbits, they are correct to $\mathcal{O}\left(e_c^2\right)$ in the eccentricity $e_c$. For all the available baselines, we compute \eqref{eqn:GGNGW} for various values of $\tau$. To maximally distinguish the GW from the GGN, we take $\tau$ ranging all the way to $\approx\frac{\pi}{\omega}$, that is all the way to half the gravitational wave period. To accurately capture the physics, we use the integrated position change from the acceleration in \eqref{eqn:minisigdeltaT}. 

First, we considered the case of two deputies and the central satellite that are all in the same circular orbit, with their positions located at a phase offset from the central satellite. In this case, the system is far too symmetric and there is an enormous SNR penalty to pay to distinguish the gravitational wave from the gravity gradient.  Second, we considered the case of two deputies that are in elliptical orbits around a central satellite that is in a circular orbit. The analysis showed that while the gravitational wave can be distinguished from the GGN, there was still a severe SNR penalty  $\sim 10^{-3}$ (on top of the velocity suppression) that had to be paid to distinguish $h$ from the GGN. We then considered the case where the central satellite is in an elliptical orbit, with the deputies also being in elliptical orbits.  In this case, for gravitational wave frequencies that are comparable to or smaller than the orbital frequency (the case that is of most interest to this paper), the SNR penalty is $\mathcal{O}\left(1\right)$. This penalty is not a serious issue for optical metrology - there is considerable room to enhance \eqref{eqn:SNRCon}. However, the SNR penalty implies that the proof mass needs to be suitably more stable, requiring more aggressive parameters on the atom interferometer configuration used to stabilize the mass. It can also be checked that with one additional satellite that is also in an elliptical orbit around the central satellite, this $\mathcal{O}\left(1\right)$ SNR penalty disappears. 

For gravitational wave frequencies higher than the orbital frequency, the SNR penalty depends on the lifetime of the source. Specifically, in order to distinguish the GW from the gravity gradient, the constellation has to physically traverse the GGN field. When the period of the gravitational wave is shorter than the orbital frequency, the distance traveled by the satellites in a gravitational wave period is not as long, making signal extraction harder - the SNR penalty can become $\mathcal{O}\left(10\right)$ or more in this case. This is partially offset by the fact the GGN background is a rapid function of the frequency - at frequencies $\sim 10^{-7}$ Hz, it is a 100 times smaller than at $\sim 10^{-8}$ Hz and falls exponentially above that. Thus, one does not have to cancel GGN at the same order. Moreover, the real SNR limit is set by the performance of the proof mass - at higher frequencies, the intrinsic acceleration of the gravitational wave is higher, making proof mass stability an easier challenge from the metrological perspective. Further, the SNR penalty decreases if the observation time is increased to the orbital period since this allows the satellites to traverse a sufficiently large distance. The limit on the observation time is set by the lifetime of the source - since most sources will survive several orbital periods before merger, extending the observation time to a few periods of the gravitational wave is not a serious limitation. A more systematic analysis of the trade-offs in this part of frequency space is warranted - but this is beyond the scope of our present work.  We checked the robustness of our simulation with a number of additional checks. We added additional satellites to provide more baselines and we also used orbits that were correct to $\mathcal{O}\left(e_c^3\right)$ in the eccentricity. Neither of these changed the SNR penalty, which remained $\sim v$ as expected. 

We should also worry about the fact that there is a relativistic time delay in the gravitational gradient field arising from the fact that the motion of the asteroids will impact different satellites at different times due to the causal Green's function that changes the local gravitational field. Time delay interferometery sequences \cite{Estabrook:2000ef} can  be performed on each baseline $jk$ to cancel the leading order ($\sim \omega L$) relativistic lag. This is because the relativistic time lag in the gravity gradient can be expanded as a Taylor series in time around the instantaneous acceleration - the first order relativistic lag is thus a jerk around the instantaneous acceleration. Arbitrary independent jerks at $j$ and $k$ can be canceled using the following protocol. Assume that the satellites $j$ and $k$ are continuously sending light from one to the other. When the light from the emitter arrives at the receiver, the phase of the light from the emitter is mixed with a local oscillator onboard the receiver. Now for the baseline $jk$, $k$ measures the phase of the light from $j$ at time $T$ and $j$ measures the phase of the light that arrives at time $T + L$ where $L$ is the separation between $j$ and $k$. It can be checked that the average of these two measurements is independent of the jerk and retains the acceleration signal. This average can then be used to calculate \eqref{eqn:GGNGW}. Note that terms $\mathcal{O}\left(v  \, \omega L\right)$ and $\mathcal{O}\left(\omega L\right)^2$ are too small for the sensitivities, frequencies and length separations of interest.  

\subsubsection{Caveats and Additional Opportunities }
The above analysis was done by assuming that the gravitational wave and the gravity gradient were at the same frequency $\omega$, with the satellite orbits being taken at a different, but known, frequency $\omega_o$. This analysis, while supportive of the ability to distinguish the higher order gravity gradient noise from the gravitational wave, is also incomplete. In principle, we should put the GGN noise over a broadband of frequencies and ask if the signal can be extracted from such broadband noise. This is in fact important since the modulation of the orbit will mix GGN noise at sidebands around the orbital frequency and there can be spectral leakage of the broadband GGN noise into the frequency band of interest. A complete signal processing analysis of this kind is beyond the scope of our present work and we leave it for future analysis. However, we note that for frequencies $\omega \ll \omega_o$, which is appropriate for gravitational wave frequencies $\sim 10$ nHz, our analysis should largely hold - the GGN and signal should then be clustered around sidebands of the harmonics of the carrier frequency $\omega_o$. Similarly, when $\omega \gg \omega_o$, the orbital dynamics sits in a well defined sideband around the signal (and GGN) frequency $\omega$. When $\omega \sim \omega_o$, the analysis is more involved. Since this is also a relevant frequency range, a more careful analysis is warranted.  

Our analysis also focused on orbits that were centered around the Sun. An additional possibility is to consider a primary satellite that is in orbit around the Sun, with other satellites orbiting Mars. For example, one may consider another satellite that is in a Mars orbit at a distance $\sim 10^{5}$ km from Mars, giving it an orbital period $\sim 10^6$ s. The velocity of such a satellite is $\sim$ km/s, comparable to the velocities considered in the solar orbits above. In such a configuration, one can use the signal extraction procedure described in sub-section \ref{subsec:algebra}. In this setup, the gravitomagnetic acceleration will modulate at $\mu$Hz frequencies due to the velocity of the satellite, while the GGN will modulate at the signal frequency and the orbital period of the satellites around the Sun, both of which are smaller than the velocity modulation frequency $\mu$Hz. Now the actual asteroid GGN noise at $\mu$Hz is negligible and thus one can cleanly extract the gravitomagnetic signal from the GGN. It is preferable to place such an orbit around Mars as opposed to the earth since vibrations of the planet surface is a source of tidal gravity gradient noise (see the following sub-section \ref{subsec:sunandearth}). Given the enormous amount of water moved by tides on the earth, it is likely that there is significant gravity gradient noise from the Earth. The significantly quieter surface of Mars may thus make Martian orbits more amenable to signal extraction. Note that the gravity gradient noise of Mars can also be compensated by placing additional satellites in orbit around Mars and using the known transfer function of the orbit to cancel the common source of noise. An additional advantage of such an orbit is also the fact that the distance between the primary satellite and the deputies can now be an AU, boosting the gravitational wave signal. This would help with systematic sources of noise associated with the proof mass. We leave a detailed study of this possibility for future work.

\subsection{Solar and Terrestrial Noise}
\label{subsec:sunandearth}
The radius of convergence of the Taylor series used to define the gravitational gradient potential $\Psi$ is the distance from the point of expansion to the nearest significant mass of concern. At 1 AU orbit, the Earth and the Sun are all around $\sim 1$ AU from the satellites, which is also nearly the distance to the asteroid belt. Thus gravitational effects from these bodies are also part of the same GGN expansion - that is the co-efficients $\Psi_{xx}$ etc capture the gravitational field of the Solar system in the region of the satellite constellation. While \cite{Fedderke:2020yfy} established asteroids as a major source of gravity gradient noise for low frequency gravitational wave detection, it did not establish that this was the only source of gravity gradient noise. The intent of this paper is stronger - we aim to demonstrate that gravitomagnetism enables sufficient rejection of gravity gradient noise so that proof masses in the inner solar system can be used for low frequency gravitational wave detection. 

The gravitomagnetic accelerations of interest are: 

\begin{equation}
a_{GW} \sim h L \omega^2 v \approx 10^{-21} \frac{\text{m}}{\text{s}^2\sqrt{\text{Hz}}} \left(\frac{L}{10^7\text{ km}} \right)^2 \left(\frac{h}{10^{-11}\sqrt{\text{Hz}}^{-1}} \right)\left( \frac{\omega}{\left(2 \pi \times 10^{-8} \text{ Hz}\right)} \right)^2
\label{eqn:netaccsig}
\end{equation} 
leading to a net displacement
\begin{equation}
\delta x_{GW} \sim \left(a_{GW} \frac{1}{f_{GW}^2} \right) \sqrt{f_{GW}} \sim 10^{-9} \text{ m}
\label{eqn:netpos}
\end{equation}
Note that $L^2$ scaling follows from $v\propto L$.

The biggest source of noise from the Earth comes from tidal activity which moves an enormous amount of water over the radius $R_E \sim 6000$ km of the earth by displacements $\delta x \sim $ m. The base gravito-electric acceleration from this motion at $\sim 1 \text{ AU}$ distances comes from the quadrupole acceleration: 

\begin{equation}
a_{E} \sim \frac{G J L}{R^5} \approx \frac{G \rho R_E^3}{R^2} \left(\frac{R_E L \, \delta x}{R^3} \right) \sim 3 \times 10^{-25} \frac{\text{m}}{\text{s}^2}
\end{equation}
This acceleration is roughly equal to the expected signal $a_{GW} \sqrt{f}$ - but this is a massive overestimate since the tidal frequency ($f_T \sim 0.1$ mHz) is much higher than the gravitational wave frequencies of interest here. Thus the actual position noise contributed by this tidal activity is significantly suppressed by $\left(\frac{f_{GW}}{f_{T}} \right)^2$. Notice also that this is the gravito-electric contribution - the gravito-magnetic contribution is further velocity suppressed. Thus, tidal activity from the Earth is not a significant source of background. 

The Sun is also a potential source of gravity gradient noise. The dominant sources of noise arise from coronal mass ejections \cite{CME} and low frequency solar oscillations, whose amplitudes can be estimated using measurements of helio-seismology \cite{Christensen_Dalsgaard_2002}. Coronal mass ejections involve ejection of $\sim 10^{12}$ kg of material from the Sun with velocities $v_{\text{CME}}\sim$ 1000 km/s. These occur with a frequency $f_{\text{CME}}$ of a few events per day. The net gravito-electric position changed caused by these are: 

\begin{equation}
\delta x_{\text{CME}} \sim \frac{G \Delta M}{R^2} \left( \frac{R}{v_{\text{CME}}}\right)^2 \sqrt{\frac{f_{\text{CME}}}{f_\text{GW}}} \sim 10^{-9} \text{ m}
\end{equation}
This position change is at the same level as \eqref{eqn:netpos}. But notice that the gravitomagnetic acceleration from this noise is smaller by a factor of $\sim 100$ and is thus not an issue. Additionally, these are also resolvable events - the mass ejection traverses the inner solar system in a time scale $\sim 10^5$ s and is thus not in the same frequency band as the signals of interest.  

Helio-seismology has measured low frequency seismic waves with velocities $v_{\odot} \sim 0.1$ \text{m/s} with periods $\tau_{\odot}\sim 300 \text{ s}$. These are oscillations of the photosphere with a density $\rho_{\gamma}\sim 10^{-3} \text{ kg } \text{m}^{-3}$. The gravito-electric acceleration from these oscillations is from the quadrupole

\begin{equation}
a_{\odot} \sim \frac{G \rho_{\gamma} R_{\odot}^4 L \left(v_{\odot}\tau_{\odot}\right)}{R^5} \sim 3 \times 10^{-23}\,  \frac{\text{m}}{\text{s}^2}
\end{equation}
This base acceleration is about a couple of orders of magnitude larger than $a_{GW}$ - but note that since this occurs at a higher frequency, the net displacement from this is considerably suppressed. Moreover, this is a gravito-electric acceleration - the gravito-magnetic part is severely suppressed by the velocity of the wave. Theoretical models suggest that there are deeper waves in the much denser part of the Sun. These cannot be directly measured but since they move a lot more mass, they are the actual dominant source of noise. These are expected to be waves with mm/s velocities, but moving densities $\sim 150 \, \text{gm}/\text{cm}^3$. The gravitomagnetic accelerations from these are $\sim 10^{-28} \, \text{m}/\text{s}^2$, comfortably smaller than \eqref{eqn:netaccsig}. Notice that this is also at a higher frequency compared to the signal, and thus the position change from this is additionally suppressed. 

The last bit of gravitational noise to consider are accelerations from meteorites \cite{meteors}. The acceleration noise from this is dominated by close encounters - the radius of convergence of this kind of noise is set by the impact parameter and since this impact parameter is smaller than $\sim 1$ AU distance from the asteroid belt, the gravity gradient $\Psi$ that is used to model the asteroid noise cannot be used to capture the effects of meteorites. In other words, these are local sources of disturbance and they will impact one test mass more than the other, unlike a uniform tidal stretch. Using the measured/estimated flux  $\sim 100 \text{ year}^{-1} \left( 10^6 \text{ km}^2\right)^{-1} $ of kg scale objects in the inner solar system, it can be shown that the position change from these transits over the gravitational wave period is $\sim 10^{-13} \text{ m}$, considerably smaller than the signal.

\section{Proof Mass Construction}
\label{sec:proofmass}
Gravitomagnetism can distinguish gravitational waves from gravity gradient noise. But, it cannot distinguish it from vibration noise of the proof masses. This is because the signal \eqref{eqn:deltaTjk} is a difference in the integrated position of the proof mass as a result of the gravitational wave, gravity gradient and vibrations. The gravity gradient, when sourced by distant objects like the asteroid belt, can be expanded as a Taylor series.  The same set of coefficients describe the effects of the gravity gradient on a constellation of satellites. Thus, as we have seen, with enough measurements along different baselines, these terms can be cancelled to high enough order.  This strategy worked only because terms of even higher order are naturally suppressed in the Taylor series when we are within the radius of convergence of the series. These statements are not true for vibrations - the vibrations of each satellite is an independent parameter and there is no common power series that can be used to describe the vibrations of multiple satellites with the same set of coefficients\footnote{This is also true for gravity gradients arising from bodies that are within the radius of convergence of the Taylor series used for the asteroid gravity gradient. The effect of a local body on a near proof mass is substantially different from its effect on distant masses. Thus, one needs a different strategy to deal with them. Indeed, it is for this reason that we had to explicitly compute the effects of meteorites in sub-section\ref{subsec:sunandearth}}. We thus need a strategy to realize a proof mass that is sufficiently stable at the low frequencies of interest. Note that this task is even more challenging than one might naively expect - in a typical gravitational wave detector, the proof mass has to dominantly react only to gravity with an acceleration sensitivity of $\sim h L \omega^2$. In this case, the gravitomagnetic signal is a factor of $\sim v$ smaller - thus the proof mass has to be resistant to non-gravitational accelerations to an even higher degree. While a complete study of such a proof mass system is beyond the scope of this paper, we point to interesting opportunities permitted by atom interferometry to realize such a proof mass. 

In a satellite experiment, a macroscopic proof mass, such as the one demonstrated by the LISA Pathfinder mission, faces the following challenges. The proof mass, once released into the satellite's measurement chamber after launch, will be subjected to brownian noise from the random scattering of stray gas in the satellite's vacuum. It acquires unknown charges and moments from electromagnetic disturbances such as cosmic rays and as a result responds in an unknown way to electromagnetic fields. The satellite housing the proof mass itself is subject to a variety of sources of disturbance, arising for example from fluctuations in the solar intensity and wind, as well as the response of the satellite itself to its components getting charged by cosmic rays. These random motions of the satellite will gravitationally couple to the proof mass. The LISA Pathfinder mission demonstrated \cite{Armano:2018kix} that these extreme challenges can be met at the sensitivity levels required for LISA at $\sim$ mHz frequencies. While this is undoubtedly a phenomenal technological achievement, it is difficult to extend this success to lower frequencies.  This is because the position change (the ultimate measurable in a detector) caused by an acceleration $a$ scales as $a\, T^2$ - thus, when we are interested in low frequency signals, the raw magnitude of the acceleration (and thus the force that needs to be compensated) has to be considerably smaller. Given these difficulties, it is interesting to ask if other technologies could be of use at these low frequencies.

Atom interferometers may offer such a path \cite{Dimopoulos:2008sv, Graham:2012sy,Hogan:2015xla}. We build on the   ideas of  \cite{Dimopoulos:2008sv, Graham:2012sy,Hogan:2015xla} and adopt the insights of the LISA experiment to propose a hybrid concept that can be used to realize the proof masses necessary for low frequency gravitational wave detection. The basic idea is to use the atoms as an inertial reference and use the measurement of the relative acceleration between the atoms and a macroscopic proof mass to stabilize the latter. This macroscopic proof mass then acts in exactly the same way as the proof mass of LISA. The rest of the measurement scheme operates as per the techniques developed by LISA - for example, laser phase noise in the setup is canceled using  time delay interferometric techniques. Unlike \cite{Dimopoulos:2008sv, Graham:2012sy} but similar to \cite{Hogan:2015xla},  atom interferometers are always operated by high intensity local lasers.  The operation of the interferometer through a local laser  permits the baseline between the satellites to be long, while still enabling high Rabi frequency for operating the interferometer.

\subsection{Concept}
\label{subsec:AIconcept}

On each satellite, since we are measuring accelerations along two directions (assuming an orbit confined to the ecliptic), we need a proof mass whose random accelerations are controlled to \eqref{eqn:netaccsig} along the two measurement axes. Assuming an atom source that can support a shot noise $\sim \frac{10^{-5}}{\sqrt{\text{Hz}}}$ (a key design goal of the MAGIS experiment), to detect an acceleration \eqref{eqn:netaccsig}, the interrogation time needed for the interferometer is set by the phase shift: 

\begin{equation}
k_{\text{eff}} a T^2 \approx \frac{5 \times 10^{-7}}{\sqrt{\text{Hz}}}  \left( \frac{k_{\text{eff}}}{\text{eV}}\right)\left(\frac{a}{10^{-21} \frac{\text{m}}{\text{s}^2 \sqrt{\text{Hz}}}}  \right) \left( \frac{T}{10^4 \text{ s}}\right)^2
\end{equation}
Note that the distance between the satellites can be increased to around to $\sim 20 \times 10^6$ km without increasing the highest order GGN term that needs to be canceled.  Since the acceleration scales as $\propto L^2$, it can be checked that the actual phase shift with a $\sim 20$ million km baseline is in fact $\sim 10^{-5}/\sqrt{\text{Hz}}$. 

\begin{figure}[htbp]
\centering
\begin{tikzpicture}[scale=1.1]
    % Solar Radiation
    \foreach \y in {0.5, 1.5, 2.5} {
        \draw[->, thick, orange] (-1.5, \y) -- (0, \y);
    }
    \node[orange] at (-0.8, 3) {\textbf{Sun}};

    % Sun Shield
    \fill[black] (0.2, 0) rectangle (0.4, 3);
    \node[align=center] at (0.3, 3.4) {\textbf{Sun Shield} \\ (5 m$^2$)};

    % Satellite Body
    \draw[thick] (0.6, 0.5) rectangle (2.8, 2.8);
    \node at (1.55, 2.5) {\textbf{Satellite}};
    
    % Macroscopic Proof Mass
    \fill[gray!40] (1.0, 1.2) rectangle (1.6, 1.8);
    \node[align=center] at (1.3, 0.9) {\small Proof \\[-0.5ex] \small Mass};
    
    % Laser Source
    \fill[green!50!black] (2.1, 1.3) rectangle (2.4, 1.7);
    \node at (2.25, 1.1) {\small Laser};

    % Shadowed Vacuum Region
    \fill[gray!15] (3.0, 0.5) rectangle (11.5, 2.8);
    \node[gray!80!black, font=\itshape] at (7.5, 3.0) {30 \text{m} Shadowed Vacuum};

    % Retro-reflector
    \fill[blue!80!black] (11.5, 0.8) rectangle (11.7, 2.2);
    \node[align=center] at (11.6, 0.1) {Retro- \\ reflector};

    % Laser Beams
    \draw[<->, thick, green!70!black, dashed] (2.8, 1.8) -- (10.5, 1.8);
    \node[green!60!black] at (6.5, 2.0) {\small $\pi/2 - \pi - \pi/2$ sequence};

    % Atom Cloud 1 (Moving Right)
    \fill[red] (3.5, 1.8) circle (0.15);
    \draw[->, thick, red] (3.7, 1.8) -- (5.2, 1.8) ;
    \node[red, align=center] at (3.7, 2.3) {\textbf{Atoms} \\ };

\end{tikzpicture}
\caption{The hybrid atom-interferometer proof mass concept. A sun shield creates a 30-meter dark vacuum region behind the satellite. Ultra-cold atoms produced in the satellite are sent to this region. Laser pulses from the satellite, along with  their reflections from the retro-reflector, are used to execute Mach-Zender $\pi/2-\pi-\pi/2$ sequences to measure non-gravitational accelerations, which are fed back to fully stabilize the central macroscopic proof mass. We need such a setup along each measurement axis - thus, two such interferometers for each satellite.}
\label{fig:concept}
\end{figure}
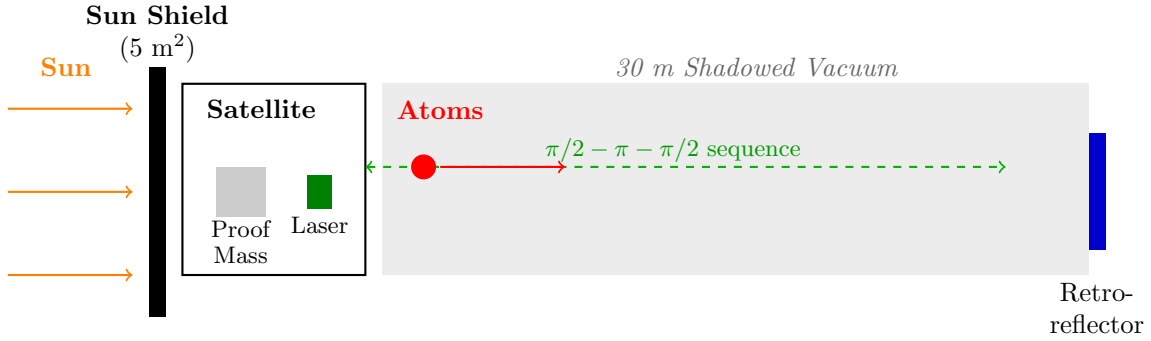

We thus need to operate the interferometer for times $T \sim 10^4$ s to sense the small acceleration \eqref{eqn:netaccsig}. This long time is serious design constraint that motivates the choices below. Following the ideas of \cite{Dimopoulos:2008sv}, we propose the concept shown in figure \ref{fig:concept}.

On each measurement axis, we have one atom interferometer. The ultra-cold atom sources and the lasers needed to operate the interferometers  will be housed in the satellite. For concreteness, we will consider interferometers operated using Rb 85/87 as opposed to clock atoms such as Sr 88 or Yb 174. The latter can also be used, but they will need stronger magnetic fields to operate. We will require that the transverse velocities of the atoms to be cooled to  $\sim$ pK, a factor of $\sim$ 50 better than presently demonstrated \cite{Kovachy:2015ygf}. Such cooling might be possible in space due to the long time available for the atom cloud to expanding, potentially enabling increased efficiency in the operation of flux preserving cooling schemes. We will also assume that the atom source can support a phase readout $\sim 10^{-5}/\sqrt{\text{Hz}}$, a key design goal of the MAGIS experiment \cite{Dimopoulos:2008sv}. The satellite will need to deploy a sun-shield of area $\sim 5 \text{ m}^2$ on each axis so that there is $\sim$ a 30 m region behind the satellites that is in total darkness, with the width of the dark shadow being $\sim \text{ m}^2$. This shield could be similar to the ultra-light shields deployed by the James Webb Space Telescope \cite{Gardner_2023}. Alternately, it could also be realized as a $\sim 30$ m long ``boom'' that extends behind the satellite, similar to the booms deployed by the DSX space-craft \cite{DSX}. At the end of this region, it is necessary to place a retro-reflecting mirror. This could potentially be realized either as part of a retractable boom or through a suitable robotic craft that ferries itself to the desired location. The operation of the interferometer will also require a stable magnetic field. This can be provided with a $\sim$ 10 Gauss magnet located on the satellite, yielding a stable $\sim 100$ nT region out to $\sim$ 30 m to provide a stable quantization axis for the atoms. 

The reason for this setup is that the atom wave-functions have to be separated in momentum space to measure the phase. Thus, at least one arm of the interferometer will move a distance $\sim$ 30 m during the course of the experiment during the interrogation time $T \sim 10^4$ s. This interferometer operation region needs to be protected from the Sun since photons that are at the resonant frequency of the atomic transitions of the atoms used to operate the interferometer will cause decoherence in time scales $\sim$ 10 s. With the sun blocked,  $T \sim 10^{4}$ s can conceivably be achieved. After the Sun, the next possible source of decoherence is the solar wind.  This is made up of particles that move with velocities  $\sim$ 400 km/s at a density  $\sim 10 \text{ cm}^3$. The scattering cross-section of these particles with the atoms is $\sim 10^{-19} \text{ m}^2$ and this  yields a lifetime $\sim 2.5 \times 10^6$ s.  The transverse size of the $\sim$ pK cooled atoms is $\sim$ 0.1 m. To eliminate coriolis backgrounds from shot-to-shot fluctuations in the thermal velocity of the atom clouds, it is desirable to servo the laser so that it remains inertial. The atoms are in free fall around the Sun and thus over $\sim 10^4$ s, they will move $\sim 0.1$ m in the transverse direction. Thus, the basic interferometer operation conditions can conceivably be satisfied. 

The interferometry operates as follows. The atoms are first cooled inside the satellite. Using the laser in the satellite and its reflected beam off the retro-reflector, the atom cloud is brought outside the satellite. These beams are then used to execute a standard $\pi/2 - \pi - \pi/2$ Mach-Zender sequence on the atom cloud.  The phase shift at the end of the interferometer sequence can be read either by fluorescent imaging or by kicking the atom clouds back to the satellite, as discussed in \cite{Dimopoulos:2008sv}. 

The atoms are ideal proof masses and respond only to gravity. In this setup, the source of this acceleration is thus the acceleration $\vec{a}_s$ experienced by the satellite and the gravitational coupling of this acceleration on the atoms.  The resultant phase shift is

\begin{equation}
    \Delta \phi_1 = \vec{k}_{\text{eff}}.\left(\vec{a}_s +\vec{I}_1\left(\vec{a}_s\right) \right)T^2
    \label{eqn:sphericalcow}
\end{equation}
In \eqref{eqn:sphericalcow}, $\vec{I}_1$ is a known transfer function that gravitationally couples the position fluctuations of the satellite (through the acceleration $\vec{a}_s$) to accelerations of the interferometer.  Given this known transfer function, the acceleration of the satellite can be obtained. For this measurement protocol to work, we assume that the retro-reflector is either rigidly attached to the satellite or that its position is fixed relative to the satellite using a separate optical interferometer. This is necessary so that the accelerations of the satellite and the retro-reflector are common, providing a single channel for this noise to enter the interferometer. The position lock between the satellite and the retro-reflector must be as good as the desired position sensitivity of the setup - thus it must be as good as \eqref{eqn:netpos}. At every interaction point, laser phase noise enters the interferometer in the combination $\delta \Phi_S - \delta \Phi_R$ where $\delta \Phi_S$ is the noise on the laser phase front from the satellite and $\delta \Phi_R$ is the noise on the phase front from the retro-reflector. Since both of these are emitted from the same laser, the noise will mostly cancel. But, due to the finite light travel time $l \sim 30$ m between the satellite and the retro-reflector, there will be an uncanceled contribution. Drift $\delta k$ in the laser frequency causes noise $\sim \delta k \, l$, which must be smaller than $10^{-5}/\sqrt{\text{Hz}}$ for $l \approx 30$ m. The laser thus needs to have a frequency stability  $\approx 100$ Hz. 

A serious source of noise arises from shot-to-shot fluctuations of the atom cloud's position and thermal velocity spread. These fluctuations in the initial position and velocity couple to the gravity gradient from the satellite and are a source of noise. The shot-to-shot acceleration noise from position fluctuations $\delta x$ scale as $\approx \frac{G\mu\delta x}{r^3 \sqrt{N_a}}$ while the fluctuation from the thermal velocity $v_{th}$ spread is $\approx \frac{G\mu\,  v_{th} T}{r^3 \sqrt{N_a}}$ where $N_a \approx 10^{10}$ is the number of atoms in the cloud. Of these, the noise from the thermal velocity is more dangerous since it accumulates with interrogation time. For a satellite with a mass $\sim 10^3$ kg and the interferometer operated at distances $r \sim 30$ m, the thermal velocity $10 \, \mu\text{m}/\text{s}$ (from a cloud at pK temperatures) causes a shot-to-shot acceleration fluctuation $\approx \mathcal{O}\left(100\right)$ worse than required. But, this can be mitigated in two ways. First, this noise scales rapidly with distance - thus, changing the interferometer operation distance to $\mathcal{O}\left(100\right)$ m suppresses this noise by a factor of $\approx 27$. Additional suppressions can likely be achieved by taking a slightly larger number of atoms, or cooling the atoms a bit more, or by operating the interferometer for a slightly shorter time. There is thus reasonable trade-space to suppress this noise. Second, it is likely that the linear dependence on $v_{th}$ can be eliminated using a more elaborate pulse sequence - for example, a sequence of the form $\pi/2 - \pi - \pi -\pi/2$ where the time between the $\pi/2 - \pi$ pulses is $T$ and the time between the two $\pi$ pulses is $T\left(1 + 2\alpha\right)$ \cite{Dimopoulos:2008sv}. It should be possible to find positive values of $\alpha$ that cancel phases that are linearly dependent on $v_{th}$ while retaining the primary sensitivity to the acceleration of the satellite. We elaborate on this possibility in future work, when we fully develop this proof mass concept. The shot-to-shot acceleration fluctuation from position fluctuations $\delta x$ of the atom cloud are below shot noise as long as these are controlled to within $\approx$ mm. 

With the laser servoed to be inertial, we can estimate the stability requirements for the local atom interferometer. This is distinct from the pointing noise we discuss below which is the error in pointing from one satellite to another. Let us assume that on-board telescopes can be used on the satellite to provide local pointing accuracy at the level of $\approx 10^{-13}/\text{s}$ averaged over the measurement bandwidth $\sim 10^{-7}$ Hz. Note that this is a factor of $10^3$ worse than the stability demonstrated by the Gravity Probe B mission. With this level of stability, the residual centrifugal accelerations out to distances $\sim 100$ m  are smaller than the required sensitivity  \eqref{eqn:netaccsig} (integrated over the measurement band). The Coriolis acceleration due to this angular jitter on a pico-kelvin cloud of atoms with  $\approx 10^{10}$ atoms (as required for the $\approx 10^{-5} \sqrt{\text{Hz}}^{-1}$ shot noise) is comparable to the required shot noise sensitivity.

\subsection{Pointing Noise}
\label{subsec:pointing}

In section \ref{sec:protocol}, we were focused on gravitational sources of noise that could be canceled using multiple baselines. We then focused on obtaining a suitable inertial mass in sub-section \ref{subsec:AIconcept}. Another major source of noise is pointing error - the satellites will need to track each other and errors in this tracking can mimic an acceleration. Interestingly, the gravitomagnetic signal provides a way to decouple the gravitational wave signal from the need to stabilize pointing. However, as we will see, this comes with the price of a SNR penalty and likely requires a four satellite constellation rather than the three satellite constellation needed to cancel the gravity gradient. 

Pointing noise enters the signal \eqref{eqn:lightdirection}. Instead of the light being sent along $\vec{V}_{\gamma}$, suppose it is sent with an error $\vec{V}_{\gamma} + \Delta \vec{V}_{\gamma}$. The resultant change in the arrival time  $\Delta^{\gamma}T_{jk}$ can be obtained from \eqref{eqn:lightdirection} and is: 

\begin{equation}
\Delta^{\gamma}T_{jk} = - \frac{\left(\vec{V}_{\gamma} - \vec{v}_{k}\right).\Delta \vec{V}_{\gamma}}{|\vec{V}_{\gamma}-\vec{v}_k|^2} dT_{jk}
\end{equation}

A pointing error $\delta \theta_j$ leads to 

\begin{equation}
\Delta \vec{V}_{\gamma} = \delta \theta_j \vec{V}^{\perp}_{\gamma} - \frac{\delta \theta_j^2}{2} \vec{V}_{\gamma}
\end{equation}
where $\vec{V}^{\perp}_{\gamma}$ is the direction perpendicular to $\vec{V}_{\gamma}$. This error can be included in \eqref{eqn:GGNGW}. Notably, this error is entirely from the pointing at the emitter $j$ - thus, the same pointing error is inherited by all the baselines $jk$, $ji$ {\it etc} that share $j$ as the emitter. This makes it possible to distinguish this signal from the gravitomagnetic acceleration. We included this effect in our simulation. We found that for a three satellite baseline, including this noise increased the SNR penalty by a factor of $\sim$ 100. However, in a four satellite constellation, the SNR penalty reduced down to $\mathcal{O}\left(1\right)$. Thus, while the cancellation of gravity gradient noise by itself does not require more than three satellites, relaxing the pointing requirement makes it likely that a fourth satellite would be needed.

Instead of an additional satellite, an alternative possibility is to use a guide star to suitably define the pointing axis. In the rotating frame of the tracking system, the error $\delta \theta$ contributes a time shift $\sim L \delta \theta^2$ - this must be smaller than $h L v$, requiring $\delta \theta \lessapprox 10^{-11} \approx \mu \text{as}$. This is comparable to the astrometric precision attained by missions such as GAIA \cite{GAIA} and is thus likely a better path to provide the desired stability. 

\section{Conclusions and Future Directions}
\label{sec:conclusions}

In this paper, we demonstrated that gravito-magnetism can cleanly distinguish gravitational waves from the Newtonian gravity gradient noise (GGN) sourced by the motion of asteroids. Although the gravito-magnetic signal is velocity-suppressed relative to its gravito-electric counterpart, the exceptionally bright astrophysical sources characteristic of the low-frequency gravitational wave spectrum make this suppressed signal potentially accessible. We established that fully reconstructing the signal while algebraically canceling the higher-order GGN is possible with a three satellite configuration, without incurring severe signal-to-noise penalties.  Executing this measurement demands a proof mass that responds exclusively to gravity at ultra-low frequencies. While technologically daunting, we highlight that ultra-cold atom interferometers can, in principle, meet these stringent inertial requirements. Our work therefore strongly motivates the continued development of satellite-based gravitational wave detectors fundamentally anchored by atom interferometry.

Another potentially interesting direction that could be explored arises from rapid developments that have occurred in cooling and controlling ultra-cold molecules. The sensitivity in the proposed setup is simply a function of the interrogation time. The real estate cost in terms of the atoms moving $\sim 30$ m in this long time occurred due to the fact that the momentum transferred to the atom imparts a velocity kick that causes the atom to move a long distance. If the field of ultra-cold molecules advances enough to be able to produce a laser cooled sample of molecules with a mass $\sim 1000$ GeV, the real estate demands on the proposed concept quickly drop - the molecules will need to traverse distances $\sim$ meters and thus the constraints of this experiment could be realized in a much more compact setup. Additional possibilities may also arise with nano-particle interferometry. One could for example consider nano-particles with a single optically addressable color center/defect. The interferometry can be achieved using Raman transitions on this defect. This will substantially increase the mass of the interfering particle, significantly decreasing the real estate requirements. A more detailed exploration of these emerging sensing platforms for this specific application could be of interest.

Intriguingly, this hybrid architecture---using a local atom interferometer to continuously stabilize a macroscopic, LISA-like proof mass---can also be deployed for gravitational wave detection in the $\mu$Hz to mHz band. Because asteroid GGN is negligible in this higher-frequency regime, detectors can target the unsuppressed gravito-electric signal, significantly relaxing the metrological demands on the atom interferometer. For instance, reaching an expected gravitational wave strain of $h_c \sim 10^{-18}$ at $1\,\mu$Hz requires an atom interferometer to operate for only $T \sim 10^3$ s over a spatial drift of $\sim 3$ m, with a modest phase sensitivity of $10^{-3}/\sqrt{\text{Hz}}$, assuming a baseline $\sim 10^6$ km. These parameters are remarkably less stringent than those required for the nHz gravito-magnetic detector discussed above. Furthermore, they are also significantly weaker than the extreme atom-sensing capabilities demanded by mid-band ($\sim 1$ Hz) detectors, where the intrinsic strain from astrophysical sources is much lower. The application of this hybrid proof mass concept to the $\mu$Hz band thus presents a highly promising avenue that warrants detailed future study.

\section*{Acknowledgments}
This work was supported by the U.S.~Department of Energy~(DOE), Office of Science, National Quantum Information Science Research Centers, Superconducting Quantum Materials and Systems Center~(SQMS) under Contract No.~DE-AC02-07CH11359.  S.R.~is supported in part by the U.S.~National Science Foundation~(NSF) under Grant No.~PHY-1818899.
S.R.~is also supported by the Simons Investigator Grant No.~827042. R.E. is supported by the John Templeton Foundation Award No. 63595, the University of Delaware Research Foundation, and NSF Grant No. PHY-2515007.

\bibliographystyle{plain}
\bibliography{refs}

\end{document}